\documentclass[aps,groupedaddress,showpacs,letterpaper,twocolumn,reprint,nofootinbib,longbibliography,pra]{revtex4-2}
\usepackage{graphicx} % standard LaTeX graphics tool
\usepackage{orcidlink}
\usepackage{amsmath} 
\usepackage{amssymb}   % AMS tex symbols
\usepackage{color}  % for coloring edit command

\newcommand{\ket}[1]{|#1\rangle}
\newcommand{\bra}[1]{\langle#1|}
\newcommand \be{\begin{equation}}
\newcommand \ee{\end{equation}}
\newcommand \bea{\begin{eqnarray}}
\newcommand \eea{\end{eqnarray}}

\begin{document}

\title{Symmetric $C_Z$ gate for ultracold neutral atoms based on counterdiabatic driving at Rydberg excitation}

\author{I. I. Beterov~\orcidlink{0000-0002-6596-6741}}
\affiliation{Rzhanov Institute of Semiconductor Physics SB RAS, 630090 Novosibirsk, Russia}
\affiliation{Novosibirsk State University, 630090 Novosibirsk, Russia}
\affiliation{Novosibirsk State Technical University, 630073 Novosibirsk, Russia}
\affiliation{Institute of Laser Physics SB RAS, 630090 Novosibirsk, Russia}

\author{K.V. Kozenko~\orcidlink{0009-0008-5652-8281}}
\affiliation{Rzhanov Institute of Semiconductor Physics SB RAS, 630090 Novosibirsk, Russia}
\affiliation{Novosibirsk State University, 630090 Novosibirsk, Russia}

\author{Peng Xu~\orcidlink{0000-0001-8477-1643}}
\affiliation{Division of Precision Measurement Physics, Wuhan Institute of Physics and Mathematics,
Innovation Academy for Precision Measurement Science and Technology, Chinese Academy of Sciences, Wuhan 430071, China, Wuhan 430071, China}

\author{I.I. Ryabtsev~\orcidlink{0000-0002-5410-2155}}
\affiliation{Rzhanov Institute of Semiconductor Physics SB RAS, 630090 Novosibirsk, Russia}
\affiliation{Novosibirsk State University, 630090 Novosibirsk, Russia}

\date{\today}

\begin{abstract} We designed a scheme for a neutral atom Rydberg blockade $C_Z$ gate based on the double sequence of adiabatic pulses applied symmetrically to both atoms and using counterdiabatic driving for Rydberg excitation. This provides a substantial reduction in the quantum gate operation time compared to previously proposed double adiabatic schemes, and makes our scheme competitive with modern time-optimal protocols for high-fidelity entangling gates with neutral atoms. Our approach creates a bridge between fully adiabatic and time-optimal gate schemes. The use of adiabatic passage reduces the sensitivity of gate fidelity to variations in laser intensity, while counterdiabatic driving provides short gate times. The intensity and phase profiles of the laser pulse acting on the atoms are described analytically depending only on the gate duration.  We demonstrated the applicability of this scheme for single-photon and two-photon schemes of Rydberg excitation in rubidium and cesium atoms, and, for the  first time, discussed the implementation of a $C_Z$ gate using three-photon excitation of rubidium atoms. In contrast to many modern $C_Z$ gate protocols, our scheme does not generate intrinsic single-qubit phase shifts, although they still appear in two-photon configuration. We also designed a numerically optimized amplitude-robust gate with an analytically defined phase profile of the laser pulse and compared its performance with the counterdiabatic gate scheme. 
  
\end{abstract}

%\pacs{03.67.Hk,32.80.-t,32.80.Qk}
\maketitle

\section{Introduction}

	Quantum computing with ultracold neutral atoms has greatly advanced in recent years. Large-scale atomic arrays containing thousands of qubits have been demonstrated~\cite{Chiu2025, Manetsch2024} and quantum algorithms with neutral atoms were successfully implemented ~\cite{Graham2022,Ebadi2022,Oliveira2025,Chinnarasu2025}. There are new promising architectures for neutral-atom quantum computing based on logical qubits~\cite{Bluvstein2024} and new technological approaches, for example, using fiber arrays~\cite{Li2024}.  The fidelity of two-qubit gates in  experiments has reached 99.7\%~\cite{Endres2025}, opening the way to quantum error correction and the design of logical qubits~\cite{Bluvstein2024,Bluvstein2025}. High-fidelity entanglement with neutral atoms is achieved using Rydberg blockade~\cite{Jaksch2000} and  carefully designed amplitude and phase profiles of laser pulses, which are used for Rydberg excitation~\cite{Levine2019, Fu2022, Evered2023,Endres2025}. The parameters of these pulses are usually obtained using numerical optimization of the gate performance~\cite{Jandura2022, Fu2022, Evered2023, Endres2025}. Optimal control theory is also used for the preparation of  desired quantum states~\cite{Bason2012}, and in particular for the design of high-fidelity gates~\cite{Chang2023}. There are also alternative gate schemes that do not rely on Rydberg blockade~\cite{Giudici2025,Ming2024}.
	
An important feature of modern high-fidelity gate protocols for neutral atoms is that they are always symmetric, as both interacting atoms are illuminated by identical laser pulses~\cite{Levine2019, Endres2025}. At the same time, the dynamics of a two-atom system depend on its initial state and the energy of interatomic interaction, which results in entanglement between the atoms. Symmetric driving by identical laser pulses has numerous advantages, including the ability to implement quantum gates in a parallel way~\cite{Evered2023} in large atomic arrays and higher entanglement fidelity, which is achieved due to the short time duration of single Rydberg excitation and the reduced spatial inhomogeneity of wide laser beams that excite the atoms into Rydberg states~\cite{Levine2019, Evered2023, Endres2025}.

In experiments, high-fidelity entanglement of two neutral atoms has been demonstrated using the Levine-Pichler gate~\cite{Levine2019} and time-optimal protocols~\cite{Jandura2022, Evered2023,Fu2022,Endres2025}. Recently, Fromonteil et al.  designed an interesting 5-pulse sequence following the refocusing technique in NMR~\cite{Fromonteil2023}. This work emphasizes remaining interest in purely analytical approaches for the implementation of high-fidelity entangling gates, when there is no need to parametrize the complex profiles of laser pulses using many parameters obtained from extensive numerical optimization.
	
In our previous works, we proposed a double-pulse adiabatic sequence which provides a $\pi$ phase shift of the two-level system when it returns to the initial state after excitation and de-excitation~\cite{Beterov2016a, Saffman2020}. This phase shift results from phase accumulation during adiabatic passage and can be understood from simple analytical formulas~\cite{Beterov2016a}. The proposal for symmetric quantum gates based on double adiabatic sequence~\cite{Saffman2020} has attracted substantial interest in recent years~\cite{Wu2021a,Evered2023, Graham2022, Wu2022,Morgado2021,Pagano2022,Jandura2023,Fu2022,Shi2022,Mohan2023,Jandura2022,Poole2025,Burgers2022,Song2024,Fromonteil2024,Dlaska2022,Pelegri2022,Robicheaux2021,Wu2021b,Li2024a,Li2021,Li2022,Li2023,Buchemmavari2024,Li2022a,Chang2023,Liu2021,Xue2024,Li2022b,Petiziol2024,Rodriguez-Blanco2023,Guo2020,Wei2022,Xu2024,Mitra2023,Hou2024,Bosch2023,Doultsinos2025,Wu2022c,Shao2020,Shi2021,Xu2024,Mohan2025,Guo2025,Sola2024,Sola2023,Bosch2025,Vybornyi2023,Delvecchio2022,Chang2023,Dalal2023}. Recently,  it has been shown that such a double-pulse adiabatic sequence can be used for entangling distant qubits in atomic ensemble~\cite{Doultsinos2025}. The main advantage of the adiabatic passage is the reduced sensitivity of gate fidelity to variations in the intensity of laser radiation. Although sophisticated numerically-optimized gate protocols with reduced sensitivity to variation in Rabi frequency (which should be kept constant during the laser pulse) have recently been developed~\cite{Jandura2023}, below we show that the adiabatic schemes are robust both to variations in the   value and the gradient of the Rabi frequency which is of interest for experimental implementation.

%%%%%%%%%%%%%%%%FIGURE%%%%%%%%%%%%%%%%
\begin{figure}[!t]
\includegraphics[width=\columnwidth]{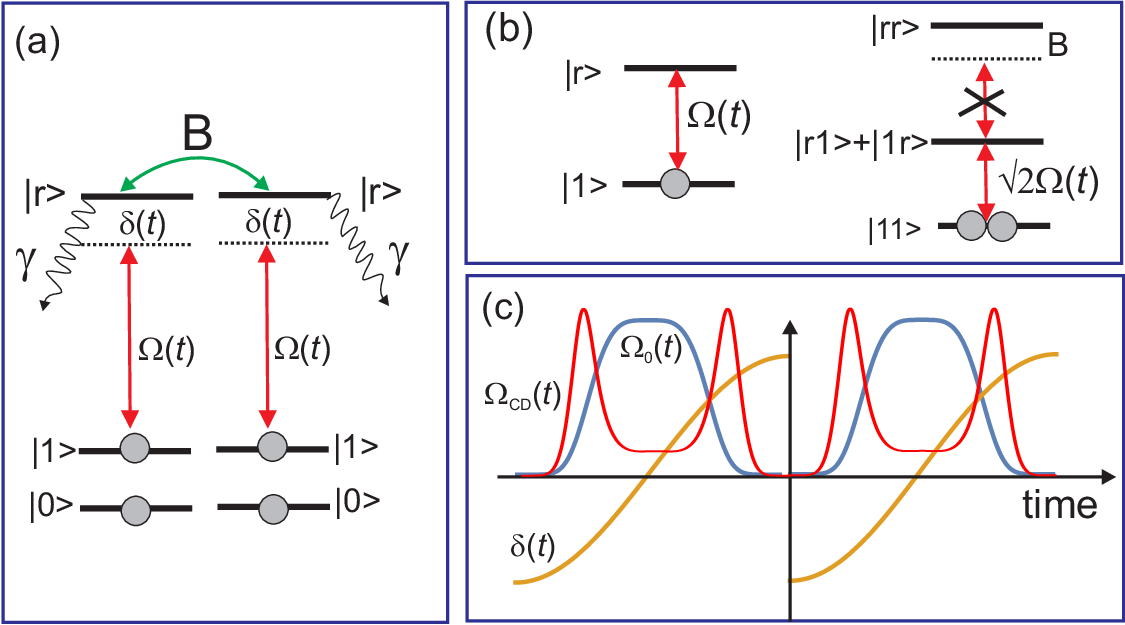}
\caption{(color online) (a) Energy level structure of two atom qubits with logical states represented by ground state sublevels $\ket{0},\ket{1}$ and Rydberg state $\ket{r}$ coupled to state $\ket{1}$ by resonant laser radiation with Rabi frequency $\Omega\left(t\right)$ and detuning  $\delta\left(t\right)$. $\sf{B}$ is the Rydberg blockade strength between atoms in states $\ket{r}$, which prevents simultaneous excitation of two Rydberg atoms. $\gamma$ is a spontaneous decay rate for Rydberg atoms.  (b) Left: Laser excitation of a single atom from ground state $\ket{1}$ to Rydberg state $\ket{r}$ by laser radiation with a single-atom Rabi frequency $\Omega\left(t\right)$.   Right: When two nearby atoms in ground state  $\ket{1}$ are illuminated by resonant laser radiation, due to the shift of two-atom $\ket{rr}$ state by out of resonance by a value  $\sf{B}$, only one atom is excited to Rydberg state. The two-atom system is effectively a two-level system with states $\ket{11}$ and $\frac{1}{\sqrt{2}}\left(\ket{r1}+\ket{1r}\right)$ and enhanced coupling $\sqrt{2}\Omega\left(t\right)$ between them. (c) A double adiabatic sequence with two smooth pulses  $\Omega_0\left(t\right)$  results in accumulation of $\pi$ phase shift after excitation and de-excitation of two-atom system prepared initially in each of the   states $\ket{01}$, $\ket{10}$, $\ket{11}$. The state $\ket{00}$ remains unaffected. The phase-shifted pulse $\Omega_{\rm{CD}}\left(t\right)$ represents a counterdiabatic term that is necessary to speed-up the gate.
}
\label{Scheme}
\end{figure}
%%%%%%%%%%%%%%%%%%%%%%%%%%%%%%%%

However, the double adiabatic passage has not yet been successfully used for the experimental implementation of high-fidelity two-qubit gates. One   possible reason is that the adiabatic protocols are intrinsically slow, which is a great disadvantage due to finite lifetimes of Rydberg atomic states~\cite{Beterov2009} and finite Rydberg blockade strengths, which limit the possibility to speed-up the gate merely by increased Rabi frequency. Therefore it is important to design quantum gate schemes that combine short gate time and moderate values of the driving Rabi frequency. The shortcut to adiabaticity~\cite{Berry2009,Hatomura2024,TORRONTEGUI2013117,Xi2010a,Morawetz2025,Latune2025,Odelin2019, An2016} is a well-known approach which allows for implementation of fast adiabatic excitation schemes by suppressing  nonadiabatic transitions to undesired states.  In the counterdiabatic driving an additional term is directly included in the Hamiltonian   to eliminate nonadiabatic coupling. This term is a function of time-dependent Rabi frequency and the detuning from the resonance in a two-level system. Counterdiabatic driving has also considerable interest in modern quantum information processing~\cite{Li2021, Finzgar2025,Romanato2025,Black2024,Wang2025}. It has been proposed to use contrdiabatic driving to suppress noise in CNOT gates~\cite{Cavalcante2025}. Recently, the preparation of Rydberg superatoms using counterdiabatic driving was studied~\cite{Yang2025}. A $C_Z$ gate for superconducting qubits using the shortcut to adiabaticity was demonstrated experimentally~\cite{Wang2019}.

Although it seems  straightforward to use the shortcut to adiabaticity to speed up two-qubit gates based on adiabatic rapid passage, the research in this area is limited. Acceleration of the double adiabatic passage for a $C_Z$ gate using the shortcut to adiabaticity was first proposed in Ref.~\cite{Bosch2023}. However, the authors of that work found that in the regime of Rydberg blockade due to $\sqrt{2}$ enhancement of the Rabi frequency~\cite{Lukin2001,Han2016,Levine2019}, the conditions for counterdiabatic driving cannot be met simultaneously for one atom and for two interacting atoms. They did not solve this problem completely, as they used a non-separable driving Hamiltonian for the system of two atoms. 

The second work that addressed this task is Ref.~\cite{Dalal2023}. The authors  proposed a more complex four-pulse protocol that allowed the gate duration to be shortened by half to $0.24\,\mu\textrm s$, compared to the initial double-pulse scheme from Ref.~\cite{Saffman2020}. However, due to finite Rydberg lifetimes, shorter and fully scalable time-optimal protocols are more promising~\cite{Endres2025}.

In this paper we show that for two-pulse adiabatic pulse sequence it is also possible to find the conditions to obtain the population and phase dynamics of the two-atom system, which is required for a $C_Z$ gate operation. This allows substantial shortening of the gate time compared to the original proposal~\cite{Saffman2020}. The overall performance of our gate is comparable to modern time-optimal protocols~\cite{Levine2019,Evered2023,Jandura2022,Jandura2023, Endres2025}. At the same time, because of the use of adiabatic passage, our gate scheme provides reduced sensitivity to variations in laser intensities.

The paper is organized as follows. In Sec.~\ref{sec.CDTheory} we discuss the theory of counterdiabatic driving and present the results of calculations for the ideal case of single-photon Rydberg excitation without spontaneous decay and with infinite blockade strength. We show that in this case, the fidelity of entanglement can be close to one. In Sec.~\ref{sec.Robustness} we analyze how the gate performance can be improved for finite blockade strength and study robustness of the gate fidelity to variations in the Rabi frequency. We also calculate the gate fidelity taking into account spontaneous decay of Rydberg states.  In Sec.~\ref{sec.TwoPhoton} we consider adiabatic rapid passage for the most common two-photon schemes of Rydberg excitation. In Sec.~\ref{sec.ThreePhoton} we show that adiabatic rapid passage with three-photon laser excitation is also feasible.  The results are summarized in Sec.~\ref{sec.Conclusion}. 

\section{Counterdiabatic driving and Rydberg blockade}
\label{sec.CDTheory}

In experiments on quantum computing with ultracold neutral atoms, well isolated hyperfine sublevels of the ground state  of alkali-metal atoms (usually rubidium and cesium) are used as qubit logical states $\ket{0}$ and $\ket{1}$, as shown in Fig.~\ref{Scheme}(a)~\cite{Saffman2016}. When the atom is excited to the Rydberg state $\ket{r}$ and returns back to the ground state, it accumulates a phase shift $\pi$ which can be used for the $C_Z$ gate~\cite{Jaksch2000,Saffman2010}. In the regime of Rydberg blockade, which is illustrated in Fig.~\ref{Scheme}(b), two atoms are simultaneously illuminated by laser radiation, which is tuned to the resonance with the transition to the  Rydberg state. Due to pairwise Rydberg interactions, the collective energy state $\ket{rr}$ acquires a large energy shift $\sf{B}$, which is known as the blockade strength~\cite{Jaksch2000}. Simultaneous laser excitation of two nearby Rydberg atoms becomes impossible, and the system of two interacting atoms effectively becomes  a two-level system with a Rabi oscillation frequency enhanced by a factor of $\sqrt{2}$ compared to the single-photon Rabi frequency $\Omega_0$~\cite{Lukin2001}. 

The scheme of the double adiabatic sequence is shown in Fig.~\ref{Scheme}(c)~\cite{Beterov2016a}. During this sequence the atoms are excited from the initial state $\ket{1}$ to the Rydberg state $\ket{r}$. Only one atom in a two-atom system can be excited due to Rydberg blockade. The second pulse returns the atoms to the initial state, and a phase shift $\pi$ is accumulated by a whole two-atom system. This is equivalent to the entangling $C_Z$ gate. The counterdiabatic driving is provided by an additional field, which can be obtained by modifying the time and phase profile of the driving laser pulse.

The $\sqrt{2}$ enhancement of the Rabi frequency in the regime of Rydberg blockade results in the inability to simultaneously return the population to the ground state for such different initial states as $\ket{10}$ and $\ket{11}$, using a single laser pulse with the same area. This can be overcome by using complex pulse shapes with numerically optimized amplitude and phase profiles~\cite{Levine2019, Jandura2022}. An alternative is adiabatic passage for Rydberg excitation~\cite{Saffman2020}. Adiabatic rapid passage is common for laser excitation of molecular levels because of the independence of transition probability from the Rabi frequency~\cite{Malinovsky2001,Brierley2012}. A number of schemes for quantum logic gates using two-photon stimulated Raman adiabatic passage (STIRAP)~\cite{Bergmann1998,Vitanov2017} and Rydberg excitation have been developed~\cite{Moller2008,Rao2014}. 

In our previous works~\cite{Beterov2013,Beterov2014} we have found that double adiabatic rapid passage returns the system to the initial state, but with a phase shift. This shift is equal to $\pi $ for two identical laser pulses and to zero if the second laser pulse is phase-shifted by $\pi$~\cite{Beterov2016a}. This allowed us to develop schemes of quantum gates with mesoscopic atomic ensembles, using adiabatic passage and Rydberg blockade~\cite{Beterov2013,Beterov2014}. In particular, we designed a scheme for a symmetric $C_Z$ gate for two atoms~\cite{Saffman2020}. Now we revise this scheme of adiabatic passage in order to introduce shortcuts to adiabaticity~\cite{Berry2009, Tang2020}, following the idea from the previous proposal for acceleration of the double adiabatic sequence~\cite{Bosch2023}.

First, we consider adiabatic rapid passage in a two-level system. The Hamiltonian for a two-level system with states $\ket{g}$ and $\ket{r}$, interacting with a chirped laser pulse (laser frequency and intensity are varied during the pulse), is written as

\be
\label{eq1}
\mathcal{H}_0\left(t\right)=\frac{\hbar }{2} \left(\begin{array}{cc} {0} & {\Omega_0 \left(t\right)} \\ {\Omega_0 \left(t\right)} & {2\delta \left(t\right)} \end{array}\right).
\ee

\noindent Here $\Omega_0 \left(t\right)$ is a time-dependent Rabi frequency and $\delta \left(t\right)$ is a time-dependent detuning from the resonance. The adiabatic evolution of the two-level system is characterized by a mixing angle $\theta \left(t\right)$~\cite{Berman2011}:
\be
\label{eq2}
\tan\left[2\theta \left(t\right)\right]=\Omega_{0} \left(t\right) / \delta \left(t\right).  
\ee

We add the counterdiabatic driving term as follows: $\mathcal{H}=\mathcal{H}_{0}+\mathcal{H}_{CD}$ with
\be
\label{eq3}
\mathcal{H}_{CD}\left(t\right)=\frac{\hbar }{2} \left(\begin{array}{cc} {0} & {-i\Omega_{\rm{CD}} \left(t\right)} \\ 
{i\Omega_{\rm{CD}}\left(t\right)} & {0} \end{array}\right).
\ee

Here 

\be
\label{eq4}
\Omega_{\rm CD}\left(t\right)=-2\dot{\theta}\left(t\right)=-\frac{\dot{\Omega_0}\left(t\right)\delta\left(t\right)-\Omega_0\left(t\right)\dot{\delta}\left(t\right)}{\Omega_0^2\left(t\right)+\delta^2\left(t\right)}.
\ee

It requires additional imaginary coupling between quantum states of the two-level system which can be achieved by a phase-shifted laser radiation field. As shown in Ref.~\cite{Bosch2023}, the imaginary part of the counterdiabatic Hamiltonian can be created using techniques based on Floquet engineering~\cite{Petiziol2018}. At the same time, methods of precise manipulation of amplitudes and phases of the driving pulses are nowadays commonly used in experiments with Rydberg atoms~\cite{Evered2023}.

Now we consider the case of Rydberg blockade which results in enhancement of the Rabi frequency.
For a two-atom system with two logical states $\ket{0}$, $\ket{1}$ and Rydberg state $\ket{r}$ the Hamiltonian has the form

\be
\label{eq5}
{\mathcal H}={\mathcal H}_{\rm 1}\otimes I + I\otimes {\mathcal H}_{\rm 2} + {\sf B}\ket{rr}\bra{rr},
\ee
\noindent where $\rm 1,2$ label each of interacting atoms, and
$$
{\mathcal H}_{1/2}=\frac{\Omega(t)}{2}\ket{r}_{1/2}\bra{1} +\delta(t)\ket{r}_{1/2}\bra{r} + \rm H.c.~.
$$
Here  $\sf{B}$  is a blockade strength. By eliminating the doubly excited Rydberg state $\ket{rr}$ we reduce this Hamiltonian to a two-level system with states $\ket{11}$ and $\frac{1}{\sqrt{2}}(\ket{1r}+\ket{r1})$ with enhanced coupling $\sqrt{2}\Omega_0\left(t\right)$. 
As the counterdiabatic term in $\mathcal{H}_{CD}$ will experience the same enhancement, the optimizing counterdiabatic term for  two simultaneously excited atoms will be written as 
\be
\label{eq6}
\Omega_{CD}^{\rm blockade}\left(t\right)=-\frac{\dot{\Omega_0}\left(t\right)\delta\left(t\right)-\Omega_0\left(t\right)\dot{\delta}\left(t\right)}{2\Omega_0^2\left(t\right)+\delta^2\left(t\right)},
\ee 
\noindent which is different from the Eq.~(\ref{eq4}). In general, the optimal adiabatic sequences with counterdiabatic driving are \textit{different} for a single-atom excitation and the Rydberg excitation of a two-atom system in a blockade regime~\cite{Bosch2023}. This is certainly an obstacle for implementing of $C_Z$ gate which requires \textit{identical} pulse sequences regardless of the initial state of two atoms. This problem was not fully addressed in Refs.~\cite{Bosch2023,Bosch2025} where a  non-separable driving Hamiltonian for a two-atom system was studied instead of the driving part of the Hamiltonian from the Eq.~(\ref{eq5}).

Although the authors of Refs.~\cite{Bosch2023,Bosch2025} clearly demonstrated advantages of counterdiabatic driving for gate performance, the Hamiltonian  obtained by them was not feasible for direct experimental implementation when the initial  ground state of the atom is not known. 

Surprisingly, we found that it is possible to obtain the parameters of laser pulses which provide the necessary time dynamics of populations and phases in \textit{both} cases. This makes the design of fast adiabatic gates a simple and robust procedure.

\begin{figure}[!t]
\includegraphics[width=\columnwidth]{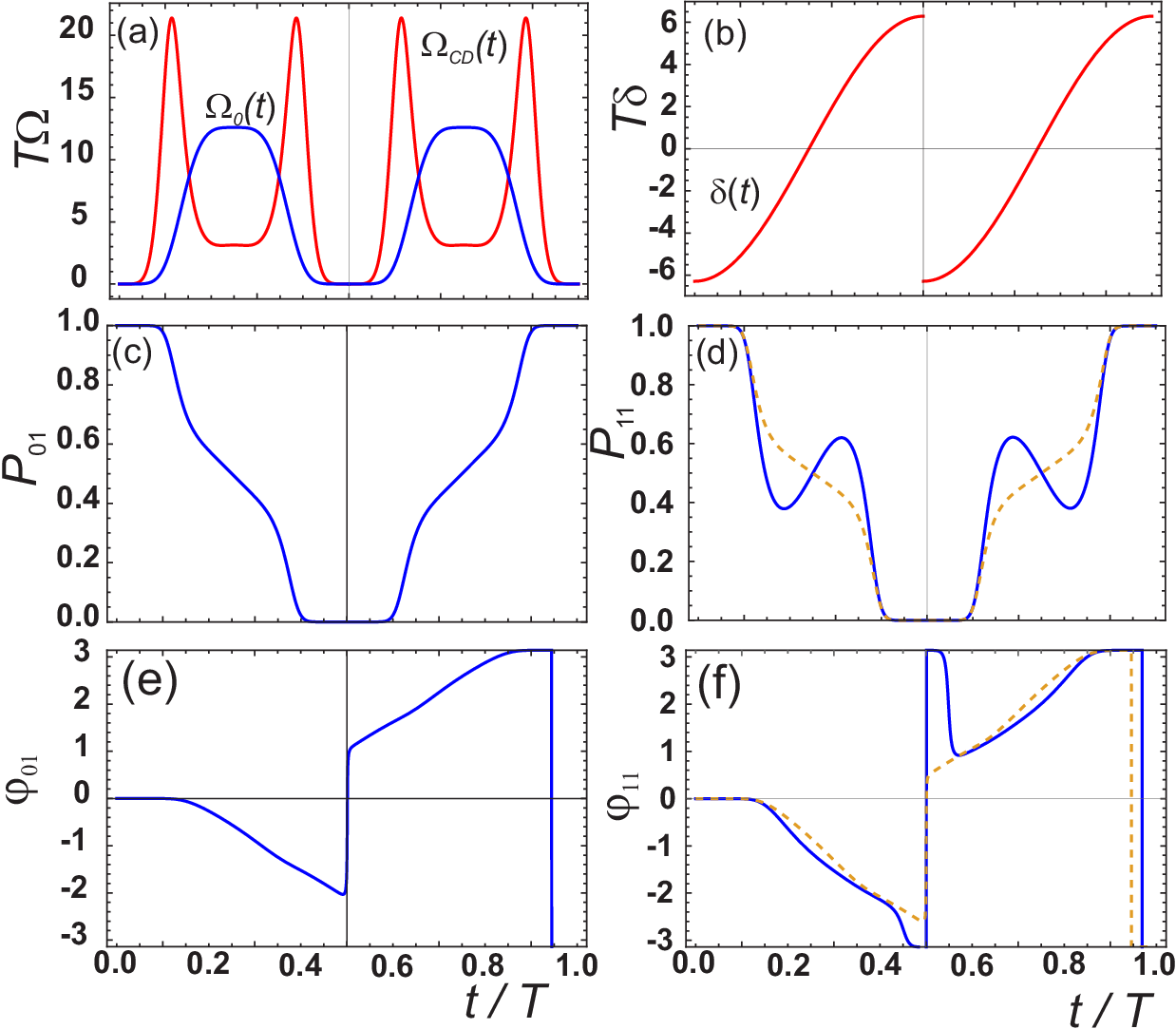}
\caption{(color online) (a)  Rabi frequency $\Omega_0\left(t\right)$ (blue) and counterdiabatic drive 
$\Omega_{CD}\left(t\right)$ (red). (b) Detuning $\delta\left(t\right)$. (c) Population $P_{01}$ of the ground state $\ket{01}$. (d)  Population $P_{11}$ of the ground state $\ket{11}$ with pulse parameters from Eq.~(\ref{eq4}) (solid line) and from Eq.~(\ref{eq6}) (dashed line). (e) Phase $\varphi_{01}$ of the ground state $\ket{01}$. (f) Phase $\varphi_{11}$ of the ground state $\ket{11}$ with pulse parameters from Eq.~(\ref{eq4}) (solid line) and from Eq.~(\ref{eq6}) (dashed line). The gate time $T$ defines Rabi frequency of the driving pulse $\Omega_{0\rm max}=4.006 \pi/T$ and  detuning $\delta_0= 2\pi/T$ for pulses defined by Eq.~(\ref{eq7}), which are centered at $t= T/4$ and $t= 3T/4$ and have $w=T/8$.}
\label{CounterdiabaticPassage}
\end{figure}

We used the following identical profiles of each laser pulse in the adiabatic sequence

\be
\label{eq7}
\begin{array}{l} 
{\Omega_0\left(t\right)=\Omega_{0\rm max} {\rm exp}\left(-\frac{(t-t_0)^4}{w^4}\right)} \\ 
{\delta\left(t\right)= \delta_0 {\rm sin} \left(\frac{2\pi(t-t_0)}{T}\right). } 
\end{array}
\ee

\noindent with $t_0$ the center of the pulse. The counterdiabatic driving term was calculated using Eq.~(\ref{eq4}) and Eq.~(\ref{eq6}).  The time dependences of the Rabi frequency $\Omega_0\left(t\right)$ and counterdiabatic drive $\Omega_{CD}\left(t\right)$ are shown in Fig.~\ref{CounterdiabaticPassage}(a). The time dependence of the detuning $\delta\left(t\right)$ is shown in  Fig.~\ref{CounterdiabaticPassage}(b).

%%%%%%%%%%%%%%%%FIGURE%%%%%%%%%%%%%%%%
%%%%%%%%%%%%%%%%FIGURE%%%%%%%%%%%%%%%%

The calculated time dependences of the populations $P_{01}$ and $P_{11}$ for the states $\ket{01}$ and $\ket{11}$ respectively are shown in Figs.~\ref{CounterdiabaticPassage} (c),(d). The time dynamics of the population are governed by different Rabi frequencies $\Omega_0 \left(t\right)$ and $\sqrt{2}\Omega_0 \left(t\right)$. Therefore, only one of the two conditions, described by either  Eq.~(\ref{eq4}) or Eq.~(\ref{eq6}) can be satisfied. We used the counterdiabatic driving term from Eq.~(\ref{eq4}) as an ansatz and  optimized two pulse parameters $\Omega_{0\rm max}$ and $\delta_0$, to identify a regime in which the system with $\sqrt{2}$ enhanced Rabi frequency also returns to the ground state, as shown in Fig.~\ref{CounterdiabaticPassage}(d). In contrast to the Levine-Pichler~\cite{Levine2019} and time-optimal~\cite{Jandura2022,Evered2023} gate schemes, our scheme does not generate intrinsic single-qubit phase shifts. This makes the optimization procedure trivial. 

For the pulse shapes defined by Eq.~(\ref{eq7}) the optimal pulse parameters for gate time $T$ are $\Omega_{\rm{0max}}=4.006 \pi/T$ and $\delta_0=2\pi/T$. For coherent laser excitation  with a constant Rabi frequency $\Omega_{\rm{0max}}$ during time $T$ the condition for a $2\pi$-pulse which returns the system to the initial state with a $\pi$ phase shift is $\Omega_{\rm{0max}}= 2\pi/T$. This establishes the ultimate limit on the minimum Rabi frequency required for gate operation. In terms of gate performance our  proposal is closer to time-optimal protocols~\cite{Jandura2022}, as it is much faster than a purely adiabatic scheme~\cite{Saffman2020}.

The detuning in Eq.~(\ref{eq7}) can be written as $\delta\left(t\right)= \delta_0 {\rm sin} \left[\delta_0(t-t_0)\right]$. All properties of laser pulses in our protocol are determined solely by the gate duration, and in the case of perfect blockade the gate is scalable for any value of $T$.

In both cases,  a $\pi$ phase shift of the ground state is accumulated, as shown in Figs.~\ref{CounterdiabaticPassage}(e),(f).  It is important to note that these parameters do not provide counteradiabatic driving at  the Rabi frequency $\sqrt{2}\Omega_0 \left(t\right)$ which would require the counterdiabatic term to follow the form defined in Eq.~(\ref{eq4}). The population and phase dynamics for true counterdiabatic driving, as defined by Eq.~(\ref{eq6}), are shown as dashed lines in Figs.~\ref{CounterdiabaticPassage}(d) and(f). Compared to true counterdiabatic driving, the necessary population dynamics  can be achieved only within a narrow window of laser pulse parameters, which resembles  on-resonance laser excitation more than adiabatic passage. This reduces the robustness of the proposed protocol, but still allows achievement of high entanglement fidelities.

\section{Analysis of gate robustness and numerical optimization}
\label{sec.Robustness}

We used the RydOpt~\cite{Rydopt2025} Python package for the analysis and optimization of gate performance. This package is designed for the simulation of the Hamiltonian dynamics of an arbitrary quantum system with analytically defined controls, taking into account spontaneous decay. The main application of the package is the numerical optimization of gate performance. It solves the Schr\"{o}dinger equation and calculates the fidelity of an arbitrary unitary gate $\mathcal{U}$ compared to a target gate  $\mathcal{U_{\rm{target}}}$ as $F=|\rm{Tr}\left({\mathcal{U}^{\dagger}\mathcal{U}_{\rm{target}}}\right)|^{2}/d^2$ where $d$ is the dimension of the  matrices. Spontaneous decay is taken into account by adding non-Hermitian terms to the Hamiltonian, which provides an accurate lower estimate of gate fidelity for errors resulting from small population leakage. The numerical optimization in the RydOpt package is based on gradient ascent.
%%%%%%%%%%%%%%%%FIGURE%%%%%%%%%%%%%%%%
\begin{figure}[!t]
\includegraphics[width=\columnwidth]{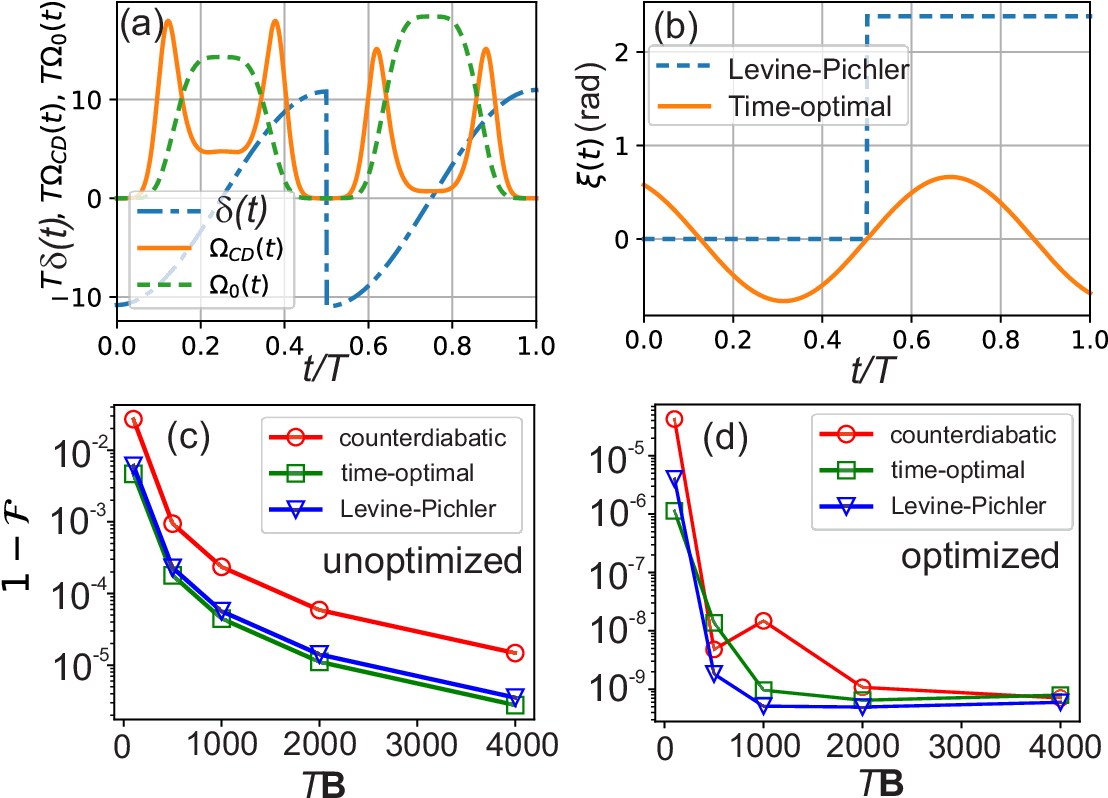}
\caption{(color online) (a) Time profiles for the counterdiabatic sequence with  independently adjusted amplitudes  $\Omega_0\left(t\right)$ (solid line) and $\Omega_{\rm{CD}}\left(t\right)$ (dashed line) of each pulse and identical shapes of detuning $\delta(t)$ (dash-dotted line). These profiles  are obtained after numerical optimization for finite blockade strength $T\sf B=100$. (b) Phase profiles for  the time-optimal gate (solid line) and the Levine-Pichler gate (dashed line).   (c) Comparison of the dependence of the gate infidelity on blockade strength $\sf B$ for unoptimized counterdiabatic gate (circles), the time-optimal gate (squares) and  the Levine-Pichler gate (triangles).  (d) Comparison of the dependence of the gate infidelity on blockade strength $\sf B$ for the numerically optimized counterdiabatic gate (circles),  the time-optimal gate (squares) and the Levine-Pichler gate (triangles).
}
\label{BlockadeStrength}
\end{figure}

The finite blockade strength $\sf{B}$ is  a well-known source of infidelities in entangling gates with neutral atoms. At moderate values of blockade strength, it results in undesirable phase shifts when the system returns to the ground state. This problem can be solved by reshaping the pulse profiles. For simplicity, in the two-pulse counterdiabatic sequence we  independently varied the amplitude of each driving and counterdiabatic pulse and the maximum detuning. An example of the non-uniform pulse profile for the counterdiabatic gate is shown in Fig.~\ref{BlockadeStrength}(a). This profile optimizes the gate fidelity for $T\sf{B}=100$. We compared the performance of the counterdiabatic gate with other modern entangling gate protocols. The pulse profile with constant Rabi frequency and detuning, and a single step in the phase of the laser pulse, is the first high-fidelity symmetric entangling gate protocol based on Rydberg blockade~\cite{Levine2019} and is known as the Levine-Pichler gate. In our numerical simulations we obtained alternative parameters of Levine-Pichler gate for $\Omega_0=1$ with the same total duration $T=8.58531$, detuning $\delta/\Omega_0=-0.3773671$ and phase shift $\xi=2.38074$. This phase profile is shown  in Fig.~\ref{BlockadeStrength}(b). The gate performance with these parameters is identical to the original proposal~\cite{Levine2019}.  It is also possible to design a gate scheme with smoothly varied phase of laser pulse. For the same gate duration, it requires sligltly smaller Rabi frequencies~\cite{Jandura2022} and is known as the time-optimal gate. The phase profile for the time-optimal gate obtained using the RydOpt package is also shown in Fig.~\ref{BlockadeStrength}(b). 

We can compare the Rabi frequencies required for different gate protocols for a certain gate duration $T=0.1\,\mu \rm s$.  The counterdiabatic scheme requires $\Omega_0^{\rm opt}=20.03\,\rm MHz$. The Levine-Pichler gate with the same duration has a Rabi frequency $\Omega_0^{\rm opt}/(2 \pi)=4.29268/(\pi T)=13.664\,\rm MHz$, Rydberg detuning $\delta_0/(2\pi)=0.377371 \Omega_0/(2 \pi)=5.16\,\rm MHz$ and phase shift $3.90242$ at the center of the laser pulse~\cite{Levine2019}.  The time-optimal scheme from Ref.~\cite{Evered2023} has the following parameters: $\Omega_0^{\rm opt}/(2 \pi)=1.215/(T)=12.15\,\rm MHz$ and the phase profile of the laser pulse is $\varphi(t)=A\rm{cos}\left[\rm{w}(t-\varphi_0\right]$  with $A/(2 \pi)=0.1122$, $\rm{w}=1.0431 \Omega_0$ and $\varphi_0=-0.7318$. The numerically optimized gate schemes illustrated in Fig.~\ref{BlockadeStrength}(b) have Rabi frequencies almost identical to the above values reported in the literature~\cite{Levine2019, Jandura2022, Evered2023}.

Figure~\ref{BlockadeStrength}(c)  shows a comparison of the dependence of gate infidelities on blockade strength for the counterdiabatic gate, Levine-Pichler gate and time-optimal gate, with parameters of all gates optimized for infinite blockade strength. For a gate duration $T=0.1\,\mu \textrm{s}$, the smallest value $T\sf{B}=100$ shown in Fig.~\ref{BlockadeStrength}(c) corresponds to $\sf{B}/(2\pi)=159~$MHz.  Without additional optimization, both the time-optimal and Levine-Pichler gates clearly outperform the counterdiabatic gate. The comparison of the same gate protocols after numerical optimization is shown in Fig.~\ref{BlockadeStrength}(d). Although the time-optical and Levine-Pichler gates still demonstrate better fidelities at lower values of $\sf{B}$, in all cases the errors resulting from finite blockade strength are below $10^{-4}$ and can be safely neglected compared to other error sources. At moderate   blockade strengths, their finite values lead mostly to phase errors instead of direct population leakage to doubly-excited Rydberg states. This is the reason why numerical optimization is so efficient for all gate schemes.

\begin{figure}[!t]
\includegraphics[width=0.65\columnwidth]{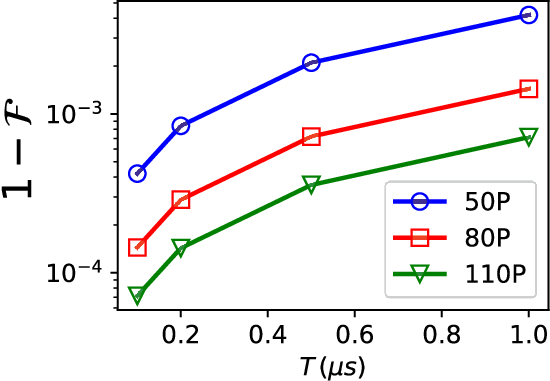}
\caption{(color online) Dependence of the infidelity on the gate duration for a single-photon configuration of laser excitation taking into account the room-temperature lifetimes of Rb 50P (circles), 80P (squares) and 110P (triangles) Rydberg states.
\label{LifetimesSinglePhoton}
}

\end{figure}

Finite room-temperature Rydberg lifetimes~\cite{Beterov2009,Vsibalic2017ARC} strongly affect two-qubit gate performance for gate fidelities above 0.999~\cite{Endres2025}.  Figure~\ref{LifetimesSinglePhoton} shows the calculated dependence of gate infidelity on gate duration $T$ for three Rb Rydberg states $50P$, $80P$, and $110P$ with room-temperature lifetimes 89~$\mu\textrm s$, 260~$\mu\textrm s$ and 525~$\mu\textrm s$, respectively. Infidelities below $10^{-4}$ can be reached with  $110P$ states for gate time duration around $0.1 \, \mu \textrm s$. For same Cs Rydberg states $50P$, $80P$ and $110P$ with room-temperature lifetimes 100~$\mu\textrm s$, 293~$\mu\rm s$ and 596~$\mu\textrm s$ the  gate infidelities are almost identical.

%%%%%%%%%%%%%%%%FIGURE%%%%%%%%%%%%%%%%

\begin{figure}[!t]
\includegraphics[width=\columnwidth]{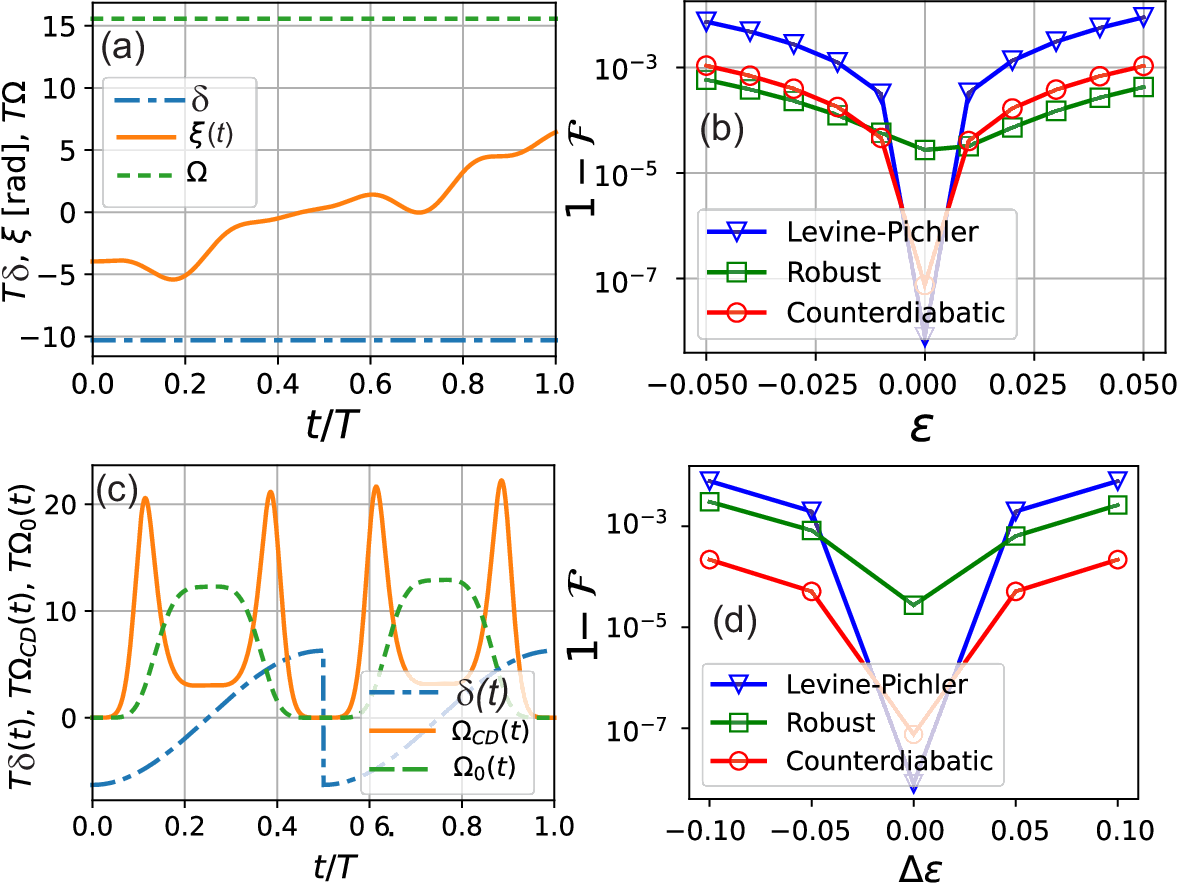}
\caption{(color online) (a) Numerically optimized pulse profiles for the amplitude-robust $C_Z$ gate with variable phase profile $\xi\left(t\right)$(solid line), constant Rabi frequency $\Omega$ (dashed line) and constant detuning $\delta$ (dash-dotted line). (b) Dependence of the gate infidelity on the variation of the Rabi frequency $\epsilon$ for the Levine-Pichler gate (circles), the amplitude-robust gate (squares) and the counterdiabatic gate (triangles). (c) Time profiles of laser pulse for the counterdiabatic gate with identical gradient of the Rabi frequency $\Delta \epsilon=0.1$ for the driving and counterdiabatic pulses. (d) Dependence of the gate infidelity on the gradient of the Rabi frequency $\Delta\epsilon$ for the Levine-Pichler gate (circles), the amplitude-robust gate (squares) and  the counterdiabatic gate (triangles). }
\label{RabiVariation}
\end{figure}

The schemes of quantum gates with increased robustness to variation of Rabi frequency were designed using numerical optimization of phase profiles of laser pulses described as piecewise functions ~\cite{Jandura2023}. As an alternative, we used chopped random basis (CRAB) initial phase profiles with an additional linear shift. Then we optimized the pulse parameters to maximize the fidelity of a $C_Z$ gate. We averaged the fidelities calculated for pulses with similar phases and durations, but with small variations of the constant Rabi frequency $\Omega_0 \to \Omega_0(1\pm\epsilon)$ with $\epsilon=0.05$. Then  we found the laser pulse profile with the maximum averaged fidelity, which is shown in Fig.~\ref{RabiVariation}(a). We found that the presence of a linear part of time-dependence of the phase  during the laser pulse is essential for our amplitude-robust scheme.

%%%%%%%%%%%%%%%%FIGURE%%%%%%%%%%%%%%%%
%%%%%%%%%%%%%%%%FIGURE%%%%%%%%%%%%%%%%

The dependence of gate fidelities on variation of Rabi frequency shown in Fig.~\ref{RabiVariation}(b) clearly demonstrates the increased robustness of our amplitude-robust gate scheme compared  to the Levine-Pichler gate. Although we have not reached the fidelities reported in Ref.~\cite{Jandura2023}, our simple optimization procedure allowed increase of robustness by an order of magnitude. The counterdiabatic gate scheme is also more robust compared to the Levine-Pichler gate, while being outperformed by  the numerically optimized amplitude-robust gate. At the same time, for $\epsilon=0$ both the Levine-Pichler and the counterdiabatic gates exibit higher fidelities.

However, in the design of the amplitude-robust gates it was assumed that the Rabi frequency remains constant during the laser pulse. We  analyzed the sensitivity of fidelity to variations in the Rabi frequency during the pulse. For this purpose, we used a linear time profile of the Rabi frequency $\Omega_0[1+\Delta\epsilon(t-T/2)/T]$ both for the driving and counterdiabatic fields, as both of them are expected to be created by phase modulation of the same laser pulse. The reshaped counterdiabatic pulse sequence with $\Delta\epsilon=0.1$ is shown in Fig.~\ref{RabiVariation}(c). In this case the counterdiabatic passage demonstrates superior robustness to the gradient of Rabi frequency, compared to both amplitude-robust and the Levine-Pichler gate, as shown in Fig.~\ref{RabiVariation}(d). Although we have not confirmed this directly, we expect a similar performance in comparison to more sophisticated amplitude-robust gate schemes from Ref.~\cite{Jandura2023}, which were also optimized for the constant value of Rabi frequency during the laser pulse.

%%%%%%%%%%%%%%%%%%%%%%%%%%%%%%%%%%%%%%%%%%%%%%%%%%%%%%%%%%%%%%%%%%%%%%%%%%%%%%%%%%%%%%%%%%%%%%%%%%%%%%%%
%===================================================================================================================
%===================================================================================================================
%===================================================================================================================

\section{Two-photon $C_Z$ gate protocol}

\begin{figure}[!t]
\includegraphics[width=\columnwidth]{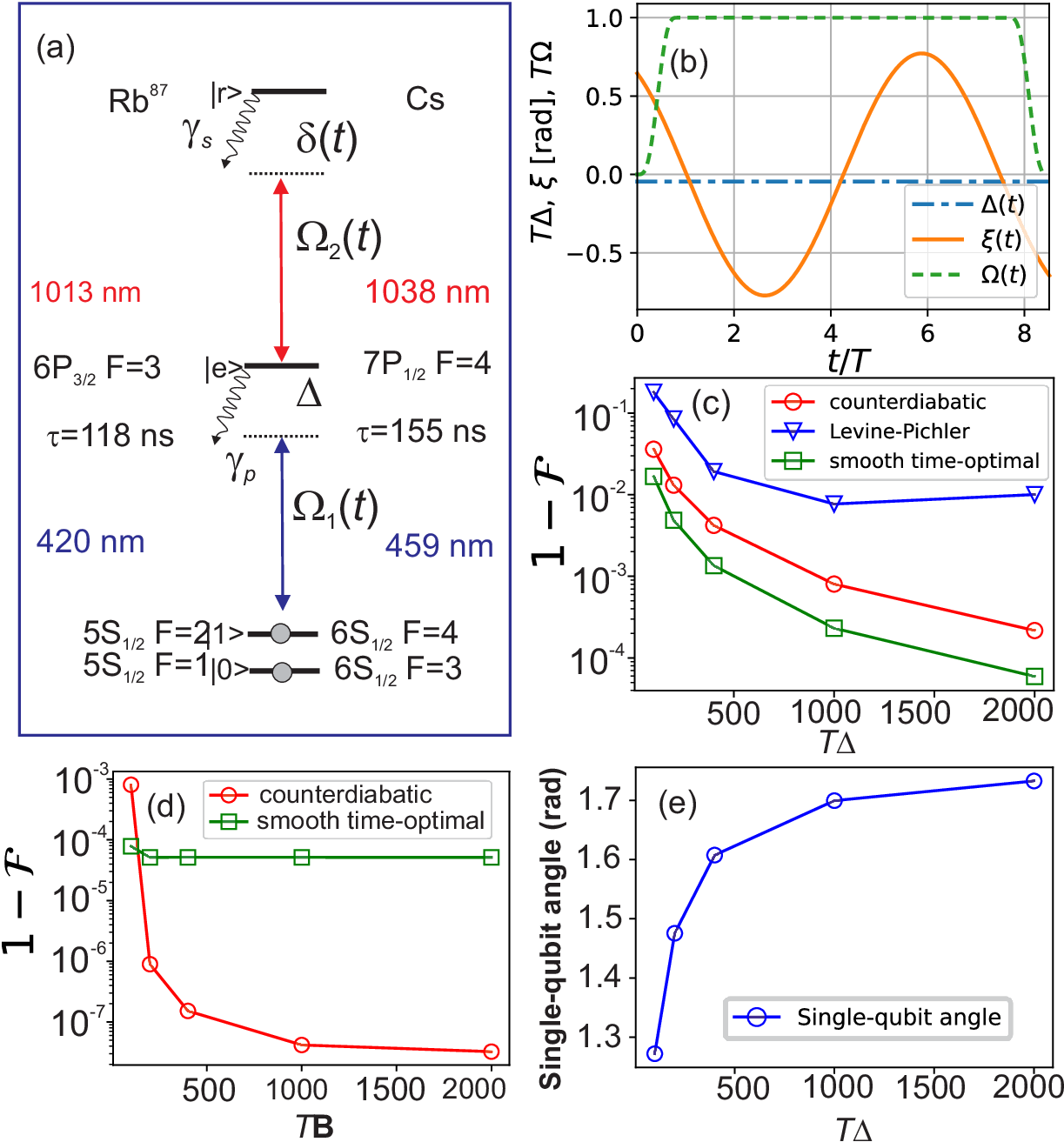}
\caption{(color online) (a) Schematic of the two-photon Rydberg excitation in Cs and Rb atoms with Rabi frequencies $\Omega_1(t), \Omega_2(t)$ via the intermediate state $\ket{p}$, with a detuning $\Delta$ from the intermediate state. (b)  Time profiles of the smooth time-optimal gate with the Rabi frequency  $\Omega\left(t\right)$ (dashed line), phase $\xi \left(t\right)$ (solid line) and constant detuning $\delta(t)$ (dashed-dotted line).  These profiles are obtained after numerical optimization for a single-photon excitation and scaled such that the maximum value of the Rabi frequency is $\Omega=1$, similarly to the parameters of the original Levine-Pichler gate.  (c) Dependence of the gate infidelity  on detuning $\Delta$ for the counterdiabatic gate (circles),  the Levine-Pichler gate (triangles) and smooth time-optimal gate (squares). (d) Dependence of the gate infideility on blockade strength $\sf{B}$ at $T\Omega_2=2000$ for the numerically optimized counterdiabatic gate (circles) and the smooth time-optimal gate (squares). (e) Dependence of the single-qubit phase shift $\phi_{01}$ of the ground state $\ket{01}$ on the detuning  $\Delta$ for the two-photon counterdiabatic gate.    }
\label{TwoPhoton}
\end{figure}

Two-photon schemes of Rydberg excitation, shown in Fig.~\ref{TwoPhoton}(a), are the most common in experiments with high-fidelity gates for ultracold neutral atoms~\cite{Levine2019,Evered2023,Maller2015}. For rubidium, the ground-state $\ket{5S}$ atoms are excited through the intermediate $\ket{6P_{3/2}}$ state to Rydberg \textit{nS} or \textit{nD} states using laser fields with 420~nm and 1013~nm wavelengths at first and second excitation steps, respectively~\cite{Levine2019}. For cesium with the $\ket{6S}$ ground state, the excitation path goes through the intermediate $\ket{7P_{1/2}}$ state using laser radiation at 459~nm and 1038~nm at first and second excitation steps, respectively~\cite{Maller2015}. From Fig.~\ref{TwoPhoton}(a) it is clear that the lifetimes of intermediate excited states for rubidium and cesium are close to each other, which ensures similar gate performance also in a two-photon configuration.

The commonly used technique for adiabatic laser excitation in three-level systems is STIRAP~\cite{Bergmann1998}. A shortcut to adiabatic passage for STIRAP requires additional coupling between first and third energy levels~\cite{Xi2010} which is challenging in ladder schemes used for Rydberg excitation. Therefore, instead, we consider a two-photon adiabatic passage, which has recently attracted interest for quantum information processing with Rydberg atoms~\cite{Pelegri2022}.

The Hamiltonian describing two-photon excitation for a single atom labeled as 1 or 2 is written similarly to our previous work~\cite{Saffman2020}
\begin{eqnarray}
{\mathcal H}_{\rm 1/2}&=&\frac{\Omega_1(t)}{2}\ket{p}_{\rm 1/2}\bra{1}+\frac{\Omega_2(t)}{2}\ket{r}_{\rm 1/2}\bra{p}\nonumber\\
&+&\Delta\ket{p}_{\rm 1/2}\bra{p} +\delta(t)\ket{r}_{\rm 1/2}\bra{r} + \rm H.c..~\nonumber
\end{eqnarray}

\noindent In order to suppress the spontaneous decay from the intermediate excited states, large detuning $\Delta$ on the order of several GHz is required~\cite{Evered2023}. In this case, the intermediate excited state can be adiabatically eliminated, and the three-level system is reduced to a two-level system with a Rabi frequency $\Omega_{\rm two-photon}=\Omega_1 \Omega_2 /2 \Delta$, where $\Omega_1$ and $\Omega_2$ are the Rabi frequencies of the first and second excitation steps, respectively, and $\Delta$ is the detuning from the intermediate excited state. When $\Omega_1  \neq \Omega_2$, there is an additional light shift of the two-photon resonance~\cite{Saffman2010}. For complex time-dependent profiles of laser pulses required for counterdiabatic driving the compensation of this light shift is not straightforward. Therefore we consider identical profiles of the Rabi frequencies $\Omega_1$ and $\Omega_2$. 
\label{sec.TwoPhoton}

We used two identically shaped complex pulse profiles 
\be
\label{eq_Rabi_twophoton}
\Omega_1\left(t\right)=\Omega_2\left(t\right)=\sqrt{2 \Delta \left(\Omega_0\left(t\right)-i\Omega_{\rm CD}\left(t\right)\right).}
\ee

\noindent with $\Omega_0\left(t\right)$ from Eq.~\ref{eq7} and  $\Omega_{\rm{CD}}\left(t\right)$ calculated using Eq.~\ref{eq6}. After adiabatic elimination of the intermediate state this gives an effective two-level Hamiltonian $\mathcal{H}$ which is similar to the  previously considered single-photon case.

%%%%%%%%%%%%%%%%FIGURE%%%%%%%%%%%%%%%%
%%%%%%%%%%%%%%%%%%%%%%%%%%%%%%%%
We compared the performance of the two-photon counterdiabatic gate  with the Levine-Pichler and time-optimal schemes. For clarity, at this point we ignored spontaneous decay. Additionally, we considered a smooth-shape time-optimal scheme~\cite{Evered2023}. Using the RydOpt package, we numerically optimized a slowly rising pulse profile for a single-photon excitation, which is shown in Fig.~\ref{TwoPhoton}(b).  From Fig.~\ref{TwoPhoton}(c) it is clear that the same smooth-shape pulse adapted for two-photon excitation without further optimization demonstrates better performance than both the Levine-Pichler and counterdiabatic schemes. However, it is possible to numerically optimize the counterdiabatic scheme both for finite values of the intermediate detuning $\Delta$ and finite blockade strength $\sf{B}$. Figure~\ref{TwoPhoton}(d) shows comparison of the numerically optimized counterdiabatic gate with the smooth-shape time-optimal gate, also optimized for finite $\sf{B}$ within a single-photon model. We did not succeed in further optimization of the two-photon smooth-shape gate for finite values of $\Delta$ using RydOpt. From Fig.~\ref{TwoPhoton}(d) it is clear that the two-photon counterdiabatic gate can be efficiently  optimized both for finite detuning from the intermediate state and for finite blockade strength.

%%%%%%%%%%%%%%%%FIGURE%%%%%%%%%%%%%%%%
\begin{figure}[!t]
\includegraphics[width=0.65\columnwidth]{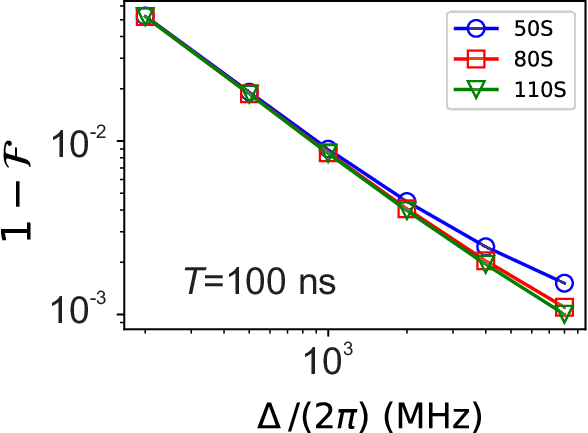}
\caption{(color online)   Dependence of the infidelity of the two-photon counterdiabatic gate  on the detuning $\Delta$ from the intermediate Rb 6P state  taking into account the room-temperature lifetimes of the Rydberg 50S (circles), 80S (squares) and 110S (triangles)  states, as well as the lifetime of the intermediate state. 
   }
\label{TwoPhotonLifetimes}
\end{figure}
%%%%%%%%%%%%%%%%%%%%%%%%%%%%%%%%

Due to the non-infinite value of $\Delta$ the adiabatic elimination is not perfect. For a two-photon counterdiabatic gate, there is an additional single-qubit phase shift, compared to the phases shown in Fig.~\ref{CounterdiabaticPassage}(e),(f) for a single-photon case. The phase shift $\phi_{01}$ accumulated after returning to the ground state for the initial $\ket{01}$ state is shown in Fig.~\ref{TwoPhoton}(d) as a function of detuning from the intermediate state. This phase shift must be corrected after the end of the gate using additional single-qubit phase gates. 

Relatively short lifetimes of the intermediate excited states strongly limit the gate performance compared to a single-photon scheme. Taking into account the decay of the intermediate state 6P\textsubscript{3/2}, commonly used for two-photon excitation of Rb atoms, with a lifetime of 118~ns, and decay of Rydberg states, we calculated the dependence of gate fidelity on $\Delta$  for a fixed gate duration \textit{T}=100~ns for different Rydberg states 50S, 80S and 110S with  lifetimes 63~$\mu \textrm s$, 205~$\mu \textrm s$, and 438~$\mu \textrm s$, respectively. The calculated infidelities are shown in Fig.~\ref{TwoPhotonLifetimes}. Only for large detunings of the order of several GHz does the increase of gate fidelity for higher Rydberg states becomes noticeable. For $\Delta/(2\pi)=8$~GHz the calculated infidelity is below 0.001.

%%%%%%%%%%%%%%%%%%%%%%%%%%%%%%%%%%%%%%%%%%%%%%%%%%%%%%%%%%%%%%%%%%%%%%%%%%%%%%%%%%%%%%%%%%%%%%%%%%%%%%%%
%===================================================================================================================
%===================================================================================================================
%===================================================================================================================
%===================================================================================================================
\section{Three-photon $C_Z$ gate protocol}
\label{sec.ThreePhoton}

%%%%%%%%%%%%%%%%FIGURE%%%%%%%%%%%%%%%%
\begin{figure}[!t]
\includegraphics[width=\columnwidth]{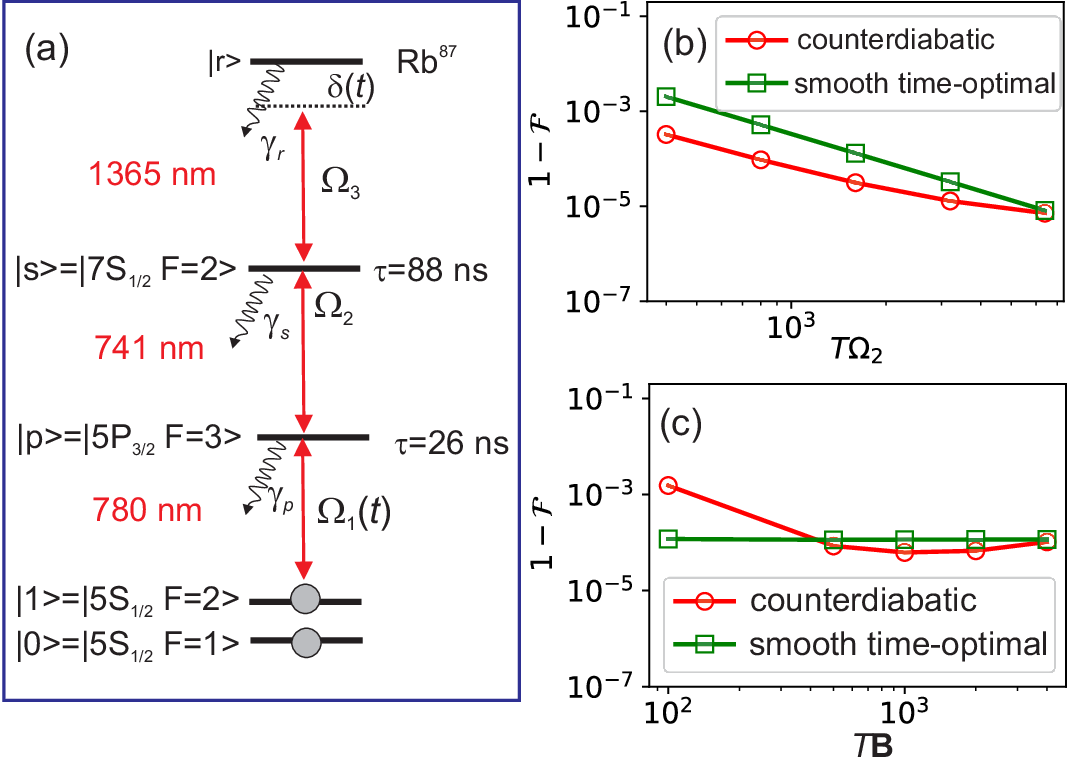}
\caption{(color online) (a) Schematic of a three-photon Rydberg excitation for Rb\textsuperscript{87} atoms. (b) Dependence of the gate infidelity on the intermediate  Rabi frequency $\Omega_2$ for the counterdiabatic gate (circles) and smooth time-optimal gate (squares).
(c) Dependence of  the gate infidelity on blockade strength $T\sf{B}$ for  the intermediate  Rabi frequency $T\Omega_2=2000$ for the counterdiabatic gate (circles) and smooth time-optimal gate (squares).
}
\label{ThreePhoton}
\end{figure}
Recently we proposed a scheme for high-fidelity individual addressing in Rydberg excitation using three-photon transitions between the ground and Rydberg state by three lasers with Rabi frequencies $\Omega_1$, $\Omega_2$, $\Omega_3$, as shown in Fig.~\ref{ThreePhoton}(a)~\cite{Bezuglov2025}. If $\Omega_2 \gg \Omega_1, \Omega_3$, even when all three lasers are tuned to the exact resonances with all three transitions, the intermediate excited states are not populated and coherent excitation of Rydberg atoms occurs, with the effective Rabi frequency $\Omega_{\rm three-photon}=\Omega_1 \Omega_3/\Omega_2$.  In this case, it is possible to compensate for the influence of the inhomogeneity of the tightly focused laser beams on the value of $\Omega_{\rm three-photon}$, which opens the way to individual addressing when performing entangling gates in atomic arrays~\cite{Bezuglov2025}. The other advantage of a three-photon scheme is the ability to \textit{completely} eliminate the Doppler shift of the three-photon resonance by geometric arrangement of three laser beams which excite Rydberg atoms~\cite{Ryabtsev2011}. Single-atom three-photon Rabi oscillations were recently experimentally demonstrated in our work~\cite{Beterov2024}.

In contrast to the two-photon scheme of laser excitation, the three-photon resonance is not shifted even  when all three Rabi frequencies are different~\cite{Bezuglov2025}. Therefore, it is easy to implement counterdiabatic driving by shaping the phase profile of only one laser beam. We apply the time-dependent detuning $\delta\left(t\right)$ to the third step of laser excitation, and the amplitude and phase profile of the first step laser pulse is modified to include the counterdiabatic driving term. 

The single-atom Hamiltonian for atoms labeled as 1 and 2 the Hamiltonian is written as
\begin{eqnarray}
{\mathcal H}_{\rm 1/2}&=&\frac{\Omega_1(t)}{2}\ket{p}_{\rm 1/2}\bra{1}+\frac{\Omega_2}{2}\ket{s}_{\rm 1/2}\bra{p}+\nonumber\\
&+&\frac{\Omega_3}{2}\ket{r}_{\rm 1/2}\bra{s}+\delta(t)\ket{r}_{\rm 1/2}\bra{r} + \rm H.c. .~\nonumber
\end{eqnarray}
Here we use constant values of $\Omega_2$ and $\Omega_3$ and the following profile of $\Omega_1 \left(t\right)$:

\be
\Omega_1 \left(t\right)=\frac{\Omega_2}{\Omega_3}\left[\Omega_0\left(t\right)-i\Omega_{\rm CD}\left(t\right)\right].
\ee

The intermediate coupling Rabi frequency $\Omega_2$ in the three-photon scheme, after adiabatic elimination of both intermediate excited states, plays the role of the detuning from the intermediate excited state in the two-photon scheme. With an increase of $\Omega_2$ the populations of the intermediate excited states become smaller due to AC Stark splitting of the resonance~\cite{Bezuglov2025}. Therefore, the gate fidelity is sensitive to the value of $\Omega_2$  even when spontaneous decay is not taken into account. Due to the absence of light-shift of the three-photon resonance, the fidelity of Rydberg excitation is tolerant to difference between $\Omega_1$ and $\Omega_3$~\cite{Bezuglov2025}. However, the best performance is achieved when $\Omega_3$ is close to the maximum value of the real part of $\Omega_1$. We have taken $\Omega_3=\sqrt{\Omega_2 \Omega_{0\rm max}}$ with $\Omega_{0\rm max}=4.006 \pi/T$.

Our simulations confirmed that a smooth-shape time-optimal scheme is more efficient for multi-photon configurations comparing to the Levine-Pichler and time-optimal gates with constant Rabi frequency. Therefore, for three-photon laser excitation we compared the dependence of fidelities on $\Omega_2$ for counterdiabatic and smooth time-optimal gate in Fig.~\ref{ThreePhoton}(b). Both schemes demontrate comparable performance with countediabatic gate being slightly better. Figure~\ref{ThreePhoton}(c) compares the dependence of gate infidelities for finite values of blockade strength $\sf{B}$ for numerically optimized smooth time-optimal and counterdiabatic gates for constant value of intermediate Rabi frequency $T\Omega_2=2000$. Here we observe better performance of smooth time-optimal gate for lower values of  $\sf{B}$ and comparable performance for higher blockade strengths.

\begin{figure}[!t]
\includegraphics[width=0.65\columnwidth]{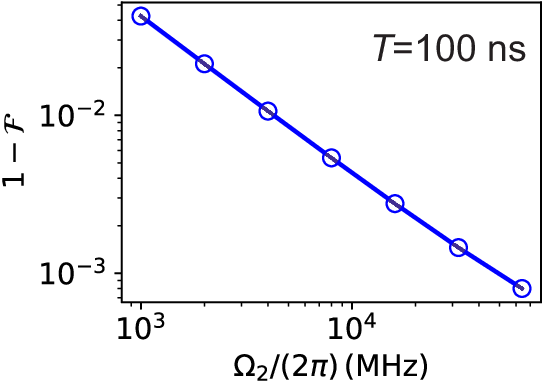}
\caption{(color online) Dependence of the infidelity of the three-photon counterdiabatic gate on the intermediate  Rabi frequency $\Omega_2$ for a gate duration $T=100$~ns,  taking into account the spontaneous decay of Rb intermediate excited 5P and 7S states and the 80P Rydberg state.
}
\label{ThreePhotonLifetimes}
\end{figure}

Spontaneous decay of intermediate excited states is the strongest limiting factor for three-photon schemes due to short lifetimes of Rb 5P and Cs 6P states which are required for all  three-photon laser excitation schemes with visible or near-infrared lasers, to the best of our knowledge. The dependence of fidelities on $\Omega_2$  for a gate duration \textit{T}=100~ns, taking into account spontaneous decay of Rb intermediate excited states and 80P Rydberg state, is shown in Fig.~\ref{ThreePhotonLifetimes}. With an increase of $\Omega_2$ the infidelity is reduced.  Figure~\ref{ThreePhotonLifetimes} clearly shows that it is necessary to substantially increase  the intermediate Rabi frequency $\Omega_2$ in order to obtain fidelities comparable to single-photon and two-photon schemes. Optimization of the three-photon gate protocols requires further efforts. Alternative schemes of laser excitation with blue and mid-IR lasers can be also considered. Similarly to a single-photon configuration, the three-photon counterdiabatic gate does not create additional single-qubit phase shifts, at least in the case when all three laser pulses are tuned to exact resonance with atomic transitions.  This makes this gate scheme interesting for further development.

%%%%%%%%%%%%%%%%FIGURE%%%%%%%%%%%%%%%%

\section{Conclusion}
\label{sec.Conclusion}

In summary, we have revised the scheme of symmetric $C_Z$ gate based on double adiabatic passage during Rydberg excitation using counterdiabatic driving. This allowed us to substantially reduce the gate operation time with a moderate increase of the Rabi frequency compared to modern time-optimal protocols, without the need for substantially larger Rydberg blockade strengths. For a case of perfect blockade, the parameters of laser pulses of our $C_Z$ gate for a general scheme of single-photon laser excitation depend on a single parameter which is gate duration time. 

The main advantages of our scheme are the following: (\textit{i}) it is almost as fast, as modern time-optimal, Levine-Pichler and amplitude-robust gate protocols, but has reduced sensitivity of the fidelity both to the value and to the gradient of Rabi frequency; (\textit{ii}) the profiles of laser pulses are described analytically without the need to use many parameters obtained from sophisticated numerical optimization; (\textit{iii}) similarly to the previously designed double adiabatic sequence, it does not create undesired single-qubit phase shifts for single-photon and three-photon Rydberg excitation.

We also obtained an analytically defined phase profile of the laser pulse for an amplitude-robust gate protocol with constant Rabi frequency and detuning.
 
The upper limit of the gate  fidelity at room temperature for the simplest single-photon excitation scheme is found to be ${\mathcal F}=0.9999$. This can be useful for future applications of quantum error correction with ultracold neutral atoms. In the calculation of this limit we have taken into account only the finite blockade strength and lifetimes of Rydberg states. These limitations are intrinsic and unavoidable, in contrast to technical imperfections of the experiment, including laser phase noise and the finite temperature of the atoms. The account of the technical factors to the gate error budget should be calculated separately, and it strongly depends on the state of art of the experimental technology. In the previous work~\cite{Saffman2020} the validity of our physical model was discussed in detail. 

We also applied our gate scheme to two-photon and three-photon configurations of laser excitation. The gate fidelity in these schemes is affected by the finite lifetimes of the intermediate excited states and non-adiabatic transitions, which limit the applicabilty of adiabatic elimination of the intermediate levels. However, we have shown that  high gate fidelities can be achieved in these more complex configurations. 

The potential advantages of three-photon laser excitation are the ability to completely compensate for the Doppler shift of the resonance at Rydberg excitation~\cite{Ryabtsev2011} and the possibility of individual addressing in atomic arrays~\cite{Bezuglov2025}. In our calculations, both single-photon and two-photon laser excitation schemes slightly outperform the three-photon excitation in terms of overall fidelity. However, the flexibility of three-photon excitation, due to the absence of light shifts and additional undesired single-qubit phase shifts (which are typical for two-photon excitation schemes), is clearly seen from the calculations. 

\appendix

\section{Numerical optimization of the counterdiabatic term for finite blockade strength}

The counterdiabatic term described by Eq.~\ref{eq4} is a function of detuning $\delta\left(t\right)$ and Rabi frequency $\Omega_0\left(t\right)$ which are defined by Eq.~\ref{eq7}. 
We re-define the pulse shapes with each pulse labeled as $i$ with $i=1,2$ as functions of three parameters:
\be
\label{eq7a}
\begin{array}{l} 
{\Omega_0\left(t,t_i,\Omega_{0i}\right)=\Omega_{0i} {\rm exp}\left(-\frac{(t-t_i)^4}{w^4}\right)} \\ 
{\delta\left(t,t_i, \delta_{0i}\right)= \delta_{0i} {\rm sin} \left(\frac{2\pi(t-t_i)}{T}\right). } 
\end{array}
\ee

We found that for numerical optimization of the counterdiabatic pulse shape to compensate for the finite blockade shift the best result is provided if we use the following   shape of \textit{i}\textsuperscript{th} pulse
\begin{eqnarray}
\label{eq4a}
\Omega_{\textrm CD}\left(t\right)&=& \\\nonumber
=  - &\frac{\dot{\Omega_0}\left(t,t_i,\Omega_{0i}\right)\delta\left(t,t_i,\delta_{01}\right)-\Omega_0\left(t,t_i,\Omega_{01}\right)\dot{\delta}\left(t,t_i,\delta_{0i}\right)}{\Omega_0^2\left(t,t_i,\Omega_{01}\right)+\delta^2\left(t,t_i,\delta_{01}\right)}.& 
\end{eqnarray}

It means that the pulse shapes in the denominator in Eq.~\ref{eq4a} are similar for both pulses 1 and 2. The amplitudes $\Omega_{0i}$ in Eqs.~\ref{eq7a} and ~\ref{eq4a} are varied independently.

\section{Numerically optimized amplitude-robust gate with analytically defined phase profile}

Using RydOpt~\cite{Rydopt2025} package we optimized  numerically the parameters of the amplitude-robust $C_Z$ gate for constant Rabi frequency $\Omega=1$, variable detuning and phase profile of the laser pulse defined by the following ansatz with $N=4$:

\begin{eqnarray}
\xi\left(t\right)&=&c_1(t-T/2)+\\
&+&\sum_{n=1}^{N}\alpha_n \textrm{sin}\left[\frac{2\pi}{T}n\left(1+\frac{1}{2}\textrm{tanh}(A_n)\left(t-T/2\right)\right)\right]+\nonumber\\
&+&\sum_{n=1}^{N}\beta_n \textrm{cos}\left[\frac{2\pi}{T}n\left(1+\frac{1}{2}\textrm{tanh}(B_n)\left(t-T/2\right)\right)\right].\nonumber
\end{eqnarray}

We averaged the fidelity for three close values of Rabi frequency and searched for the gate parameters with the largest averaged fidelity starting from random initial parameters.
The optimization procedure used random initial parameters in the range $[-1,1]$ for each parameter $A$, $\alpha$, $B$, $\beta$ of the phase profile and the initial gate duration $T$ in the range $[7,20]$. It resulted in the optimized gate duration $T=15.56$ and detuning $\delta=-0.66202$. The linear chirp of the phase profile is $c_1=0.5199$. The obtained values of the parameters A,B, $\alpha$, $\beta$ are listed in the Table~\ref{Table1}.
The time profile illustrated in Fig.~\ref{RabiVariation}(a) is rescaled for $T=1$. Random initial configurations can result in different output phase profiles. However, most of them demonstrate similar performance.

\begin{table}

\caption{\label{Table1} Numerically optimized parameters of the amplitude-robust phase profile $\xi\left(t\right)$}
\begin{tabular*}{\columnwidth}{@{\extracolsep{\fill}}|c|c|c|c|c|} \hline 
$n$&$A_n$ & $\alpha_n$ &  $B_n$  & $\beta_n $\\ \hline 
1& -0.2527502 & $0.8859972$ &   -1.17347235  & $-0.08802893 $\\ \hline 
2& -0.327713 & $-1.00445164$ &   0.10583959  & $1.20175343 $\\ \hline 
3& 0.62878523 & $0.65908966$ &   -0.25679209  & $-0.56651597 $\\ \hline 
4& 1.10736928 & $-0.32141791$ &   0.14352779  & $-0.2186016 $\\ \hline 
\end{tabular*}
\end{table}

\begin{acknowledgments}
This work was supported by the Russian Science Foundation Grant No. 23-42-00031 https://rscf.ru/en/project/23-42-00031/ in part of design of the counterdiabatic gate protocol. The numerical optimization of gate protocols was supported by the Foundation for the Advancement of Theoretical Physics and Mathematics "BASIS". P.X. acknowledges financial support from the National
Key Research and Development Program of China under 651 Grant No. 2021YFA1402001 and the National Natural Science Foundation of China under Grant No. 12261131507.
\end{acknowledgments}

\section*{data availability}
The data that support the findings of this article are openly available \url{https://github.com/beterov/Counterdiabatic-Rydberg-CZ-gate}
\bibliographystyle{apsrev4-2}
\bibliography{JCbib}{}

\end{document}